\documentclass[10pt,conference,compsocconf,letterpaper]{IEEEtran} 

\usepackage{times}
\usepackage[pdftex]{color, graphicx}
\usepackage{subfigure}
\usepackage{comment}
\usepackage{algorithm}
\usepackage{algorithmic}
\usepackage{mathtools}
\usepackage{amsmath}
\usepackage{multirow}

\newcommand{\fixme}[1]{{\em\bf{[FIXME: #1]}}}
\newcommand{\raj}[1]{{\color{red} *** {\textbf{Rajesh: }}\color{blue}{#1}}{\color{red} ***}}

\title{Missing data in multiplex networks: a preliminary study }

\author{
    \IEEEauthorblockN{Rajesh Sharma\IEEEauthorrefmark{1}, Matteo Magnani \IEEEauthorrefmark{2}, Danilo Montesi\IEEEauthorrefmark{1}}
    \IEEEauthorblockA{\IEEEauthorrefmark{1}University of Bologna, Italy
    \\\{rajesh.sharma, danilo.montesi\}@unibo.it}
    \IEEEauthorblockA{\IEEEauthorrefmark{2}Uppsala University, Sweden
    \\\{matteo.magnan\}@it.uu.se}
}

\begin{document}
\maketitle

\begin{abstract}
A basic problem in the analysis of social networks is missing data. When a network model does not accurately capture all the actors or relationships in the social system under study, measures computed on the network and ultimately the final outcomes of the analysis can be severely distorted. For this reason, researchers in social network analysis have characterised the impact of different types of missing data on existing network measures.
Recently a lot of attention has been devoted to the study of multiple-network systems, e.g., multiplex networks. In these systems missing data has an even more significant impact on the outcomes of the analyses. However, to the best of our knowledge, no study has focused on this problem yet.
This work is a first step in the direction of understanding the impact of missing data in multiple networks. We first discuss the main reasons for missingness in these systems, then we explore the relation between various types of missing information and their effect on network properties. We provide initial experimental evidence based on both real and synthetic data.
\end{abstract}


\section{Introduction }\label{sec:Intro}

Missing data is one of the main problems in social network analysis.
In most empirical studies data incompleteness is unavoidable: this applies to traditional studies in social sciences, with data coming from field experiments such as \cite{Roethlisberger1939}, \cite{Travers69anexperimental}, \cite{Bearman2002}, but also to studies targeting online sources like Facebook or LinkedIn, where the large amount of data that can be automatically collected does not necessarily imply completeness or prevent statistical biases. In all these cases it is fundamental to be able to estimate the impact of missing data on the outcomes of the network analysis process, otherwise the results may not accurately represent the studied social system.


In recent years, researchers have taken a step forward by considering systems of multiple interdependent networks \cite{Kivela2014}.
For example, consider the case of Facebook and LinkedIn.
These networks are connected with each other through the users having accounts on both of them.
This scenario introduces multiple opportunities and challenges.
If there is evidence that users are connected to the same individuals on the two networks, then the overall effect of missing data can be reduced: information missing from one network can be recovered from the other.
At the same time, collecting data from multiple networks can be more difficult, as each network has its own technical restrictions. As a result, new problems can emerge, including data missing in different ways depending on the network and new types of missing data, e.g., concerning identity relationships between user accounts in the different networks.

The general importance and practical relevance of the problem together with the differences and extensions with respect to the single-network case require a significant research effort. In this paper, we move a first step in this direction: we raise awareness of this problem and investigate some basic issues, more specifically new types of missing data and their effect on the main network properties.
\subsection{Delimitation}
Various imputation techniques have been proposed in the past to recover or estimate missing data \cite{Sadikov2011}, \cite{Bagrow2013}.
While no general approach can guarantee that the new values do not introduce any bias, imputation methods as well as approaches based on data removal are often used in practice. However, attempts to improve the quality of the data are orthogonal to our work, where we focus on estimating the impact of missingness and not on repairing it.
In addition, during the process of data collection some errors about nodes, edges and network attributes can be introduced due to wrong experimental settings or technical problems. 
This aspect is not covered in this work, because missing and wrong information require different treatments and can be considered two related but distinct problems.


\subsection{Contribution and Method}
In this work, we only deal with the impact of missing data on the networks' properties and we provide the following main contributions:
\begin{enumerate}
\item We classify the possible types of missing data in multiplex networks.
\item We characterise how some major single and multiplex network properties are distorted by different types and levels of missing data.
\end{enumerate}

We address the first item starting from known classes of missing data for single networks and discussing how these extend to the multidimensional case. The second contribution is obtained through simulation studies on synthetic and real data.

The rest of the paper is organised as follows. Section \ref{sec:prelim} summarises known reasons for missing data in social networks and discusses how this extends to multiplex networks. In Section \ref{sec:formalism}, we present our methodology to understand the impact of missing data on network properties in multiplex networks. We then evaluate the impact of missingness in multiplex networks in Section \ref{sec:eval}. We draw our conclusions and future extensions of our work in Section \ref{sec:concl}.

\section{Missing data in simplex networks}\label{sec:prelim}
\label{sec:prelim:Singnetw}

For single-layer networks, Kossinets \cite{Kossinets_SocNet2005} describes three main causes of missing data: network boundary specification, survey non-response, and respondent inaccuracy. As a working example for both the single- and multiplex-network cases, consider a network of employees at a University department.


{}
\begin{enumerate}

\item A first source of missing data is known as \emph{network boundary specification problem}, where it is the experimenter who restricts the network
by specifying rules for selection of nodes and relations \cite{lauman}.
In our University department example, the choice of restricting the analysis to the relations happening inside the institution is an example of boundary problem: two employees who are disconnected inside the department can both work together with a third individual from another department, which makes them closer than the available data would suggest. A second example consists in including only nodes with at least a given degree in the dataset.
This is a choice often made to analyse large networks and is related to the field of network sampling \cite{Leskovec2006,Gjoka2011,Salehi2013}.

\item A second source is \emph{survey non-response}, which can also lead to the loss of information \cite{Rumsey1993}. The non-response of a participant to a particular query in the survey can depend on other variables: for example, teenagers might be more willing to reveal their relationships than adults.
When other variables are involved, missing data is normally categorised into three main classes: missing completely at random (MCAR), at random (MAR) and not at random (MNAR). As we are not directly using these classes in this paper, we refer the reader to \cite{LittleAndRubin} for additional details.
\item Finally, \emph{respondent inaccuracy} concerns the fact that 
participants often only have a personal perception of their relations with other individuals \cite{Bernard1984,Marsden1990}. 
For example, chances are that A perceives her colleague B as a friend while the contrary is not true. 
\end{enumerate}

\section{Missing data in multiplex networks}\label{sec:Prelim:MissInML}

All the sources of missing data in single networks are also applicable to multilayer networks, but they are extended and complemented by others.
\begin{enumerate}
\item The \emph{network boundary specification problem} can be reinterpreted on multiple dimensions. In Figure~\ref{fig:miss-mpx-bound} we show a multiplex network where individuals can be connected on a \emph{work} layer or a \emph{friendship} layer. Consider the node in the \emph{work} network pointed by a black arrow. This node looks disconnected from the others if we only consider working relations inside the department boundaries. However, it can actually reach the other people in the network in multiple ways. One option is to pass through a common connection outside the department, in this case the grey node on the right. The grey node is external to the department, but on the same layer, connected using the same kind of relations (work). This  represents a case which we call as \emph{horizontally} breaking the network boundary. In a different case, the node can be connected to other people at the department through a connection on a different layer (friend) and thus representing a case which we term as case of \emph{vertical} boundary.


%
In addition, a node may have a degree of 100 in the \emph{friendship} layer and 1 in the \emph{work} layer, making it possible for different options of sampling based on degree. We call this specific kind of boundary problem \emph{multidimensional censoring}, e.g., the selection of a subset of objects from the network based on measures computed on multiple layers.

%


\begin{figure}[ht]
\begin{center}
   \subfigure[Multiple dimensions of boundaries]{\label{fig:miss-mpx-bound}\includegraphics[width=0.5\textwidth]{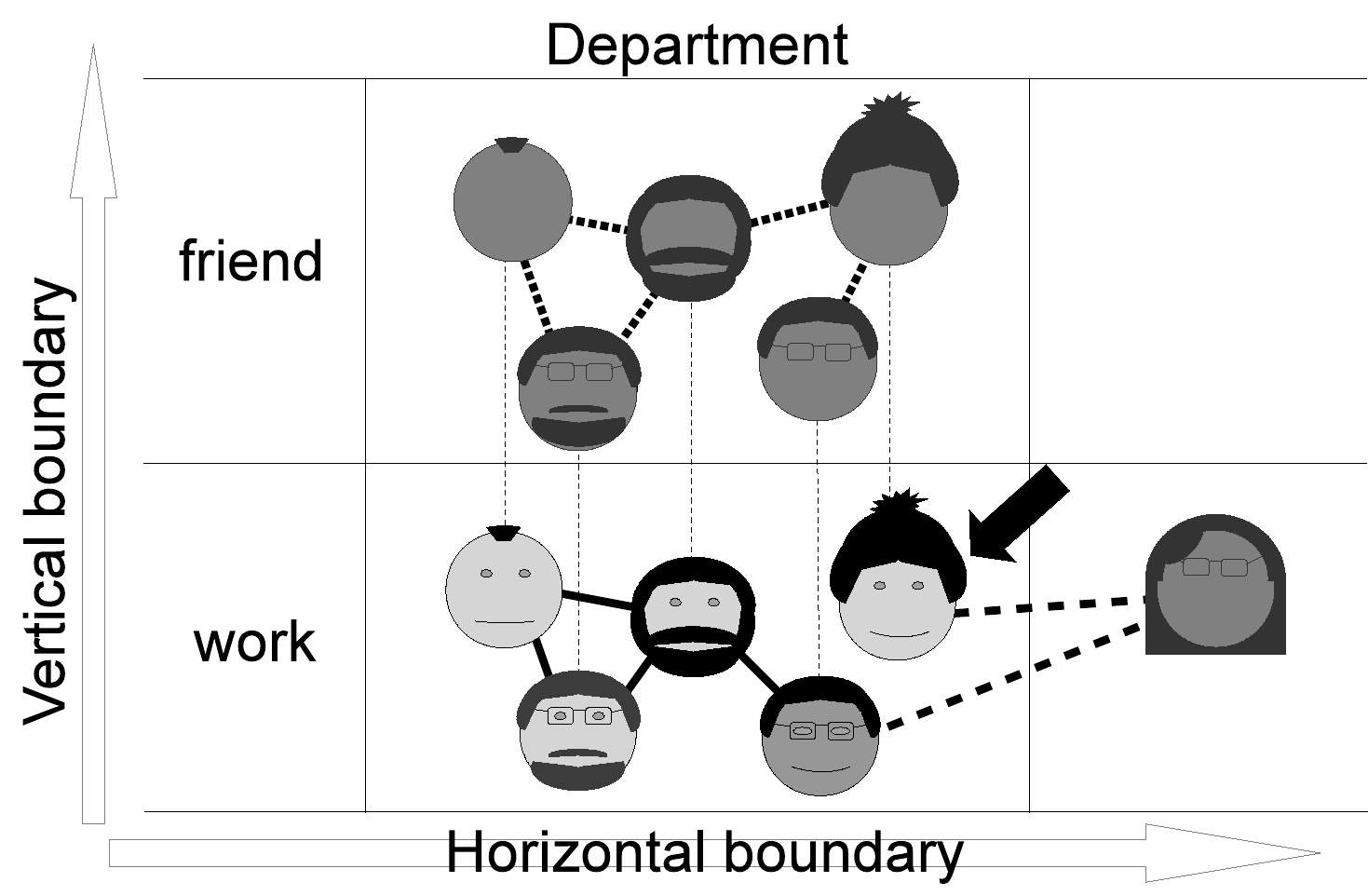}}\hspace{.5cm}
   \subfigure[Reasons for Missing data\label{fig:miss-mpx-summary}]{%
\begin{tabular}[b]{|l|}\hline
Horizontal boundary specification\\ \hline
Vertical boundary specification\\ \hline
Multidimensional node censoring\\ \hline
Survey non-response (dependent on layer)\\ \hline
Respondent inaccuracy (dependent on layer)\\ \hline
User identity resolution problem\\ \hline
\end{tabular}
}
\caption{Missing data in multiplex networks}\label{fig:miss-mpx}
\end{center}
\end{figure}

\item \emph{Survey non-response} and \emph{respondent inaccuracy} do not significantly change when we move to a multidimensional context, but both can become dependent on the layer under study. For example, people may be less willing to disclose information about their friends than their co-workers, or vice versa. As a result data can be missing from different layers in different ways. If we consider the respondent inaccuracy problem a friendship on Facebook obtained via the Facebook API can be more accurate than an offline friendship collected through a questionnaire. As we will see (in Section \ref{sec:eval}), this is relevant in practice because some layers can reduce the effect of missing data in correlated layers.



\item Multilayer network models also face specific types of missing data that do not exist in simplex networks. If we look again at Figure~\ref{fig:miss-mpx-bound}, we can see that the same nodes are present in the two layers. While this information can sometimes be easily available, e.g., when a single questionnaire is used to collect multiple relational layers, it is often the most challenging part of data to collect from online sources and must sometime rely on uncertain data integration \cite{MagnaniJDIQ10}.
We call this class \emph{user identity resolution problem}. Figure~\ref{fig:miss-mpx-summary} summarises the missing data mechanisms discussed in this section.

\end{enumerate}
\section{Methodology }\label{sec:formalism}
In this section, we first describe our methodology to understand missing data in single networks and then extend it to multiplex networks.

\subsection{Single layer networks}
A network dataset consists of i) a set of nodes ($S_N$), and ii) a set of edges ($S_E$), essentially representing a graph structure $G$($S_{N}$, $S_{E}$). In our experiments we focus on undirected networks. We consider two sets $M_N$ and $M_E$, containing respectively a list of missing nodes and edges. We then study the effect of these missing data on a set of network properties, such as diameter, average path length and clustering coefficient. These properties are indicated as $\mathrm{NP}_G$. The accuracy (or precision $P$) of the network properties for $G$ depends on how close the collected data is to the original graph. The precision of the graph properties of the incomplete graph $G'$ tends to 1 ($P$($\mathrm{NP}_{G'}$) $\rightarrow$ 1) when the cardinality of the missing data tends to zero ($\mid$ $M_N$ $\mid$ $\rightarrow$ 0 and $\mid$ $M_E$ $\mid$ $\rightarrow$ 0). The objective of our empirical study is to estimate the behaviour of $P$ at different levels of missingness by randomly removing edges and vertexes to simulate different levels of missing data. 


\subsection{Multiplex networks}\label{sec:forml:ML}
In multiplex networks, in addition to nodes and edges there is an additional \emph{layer} parameter. Each layer represents a different type of connection. Formally, a multiplex network can be represented as a graph structure $G_{ML}$=\{$N$, $E$, $L$\}, where $N$ is the set of nodes, $E \subseteq N \times N \times L$ is the set of edges, and $L$ represents the set of layers. Thus, in a multiplex network, the accuracy of the graph properties depends on three parameters ($N$, $E$, and $L$). Furthermore, the network properties in a multilayer network can be categorised in two ways. Firstly, which we termed as \emph{multiplex network properties}, a set of properties explicitly considering the layers. However, a typical way of studying a multiplex network consists in merging (some of) the layers into a single one so that traditional methods can be applied. The resultant flatten network is formed by union of all the nodes and edges of various layers of a multiplex network. In this case we use traditional single-layer metrics that we call \textit{flatten network properties}.

\subsubsection{Multiplex network properties }
In many recent works such as \cite{MatteoMuldiDimNw}, the authors have defined various multi-dimensional measures. Out of various defined measurements, in this preliminary work we focus on the exclusive relevance (or xRelevance) \cite{Berlingerio2012a}. Readers can refer to \cite{MatteoMuldiDimNw} for more properties.
For each node, xRelevance describes the importance of a particular layer in the network to exclusively reach some neighbours. \textit{xRelevance computes the fraction of neighbours directly reachable from node n following edges belonging only to a layer L}. As a consequence, removing that layer the connections with those neighbours would be lost. For example, consider a node in a multiplex network with two layers A and B, with relevance of A higher than B. If more edges are removed from A, then the relevance w.r.t A can decrease and the one w.r.t B can increase.

\subsubsection{Flatten network properties }
In the due process of flattening the multilayer networks, the resultant final flatten network very much depends on how much similar the networks in individual layers are. The Jaccard similarity between the networks of two layers $L_i$ and $L_j$ is computed as the ratio of cardinality of the intersection of the edges of two networks to the cardinality of union of edges of two networks. This notion can be extended for any number of layers. For any number of layers $l$ = 1 to L, a generalised formal representation is following,
\begin{equation} \label{EQ:Sim}
Sim(l_{i=1}^L)=\frac{\bigcap_{i=1}^L G_i}{\bigcup_{i=1}^L G_i}
\end{equation}

\begin{figure}[ht!]
  \begin{center}
   \subfigure[Diameter]{\label{fig:LC-Dia}\includegraphics[width=0.35\textwidth]{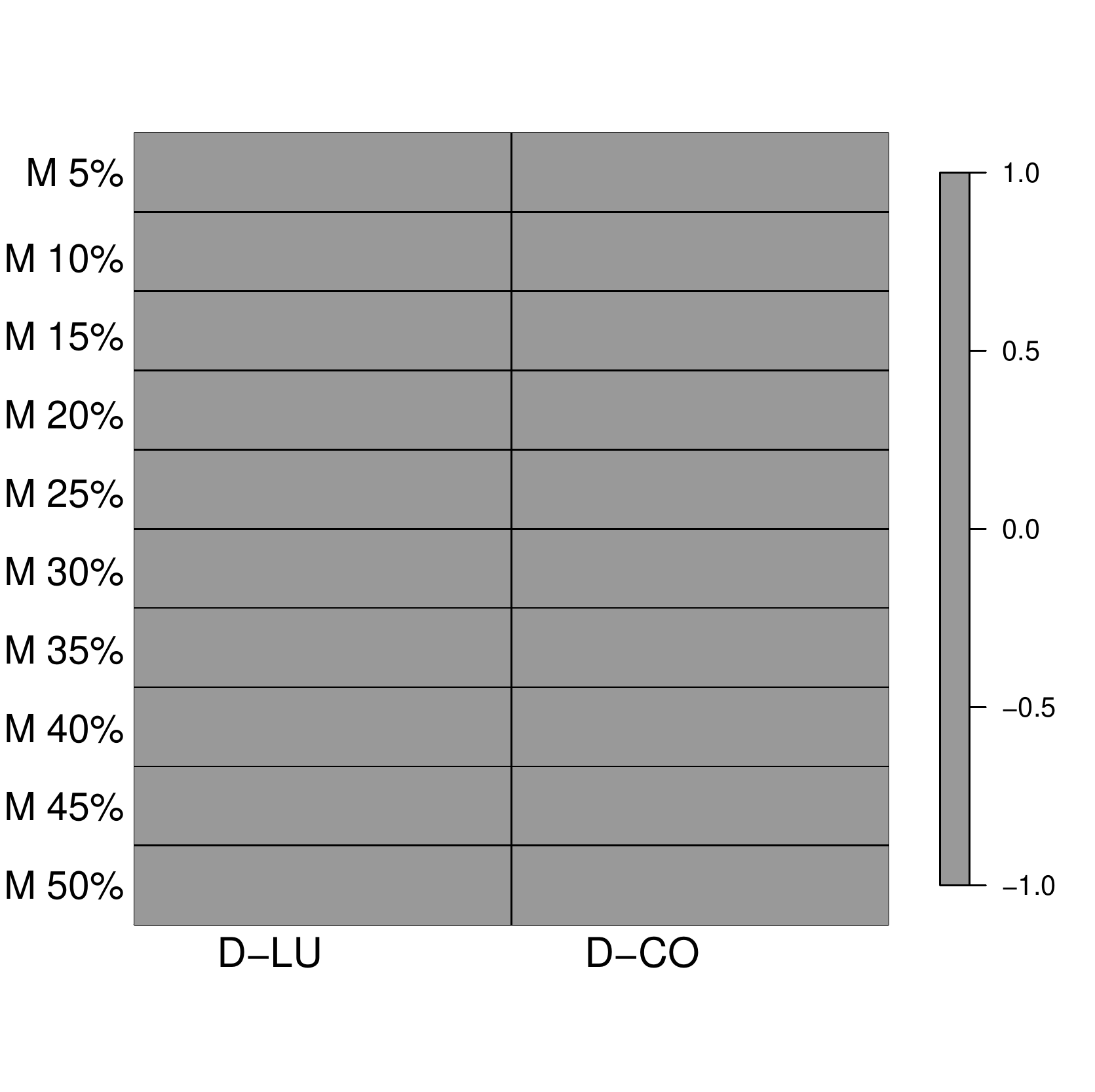}}
   \subfigure[Cluster Coeff.]{\label{fig:LC-CC}\includegraphics[width=0.35\textwidth]{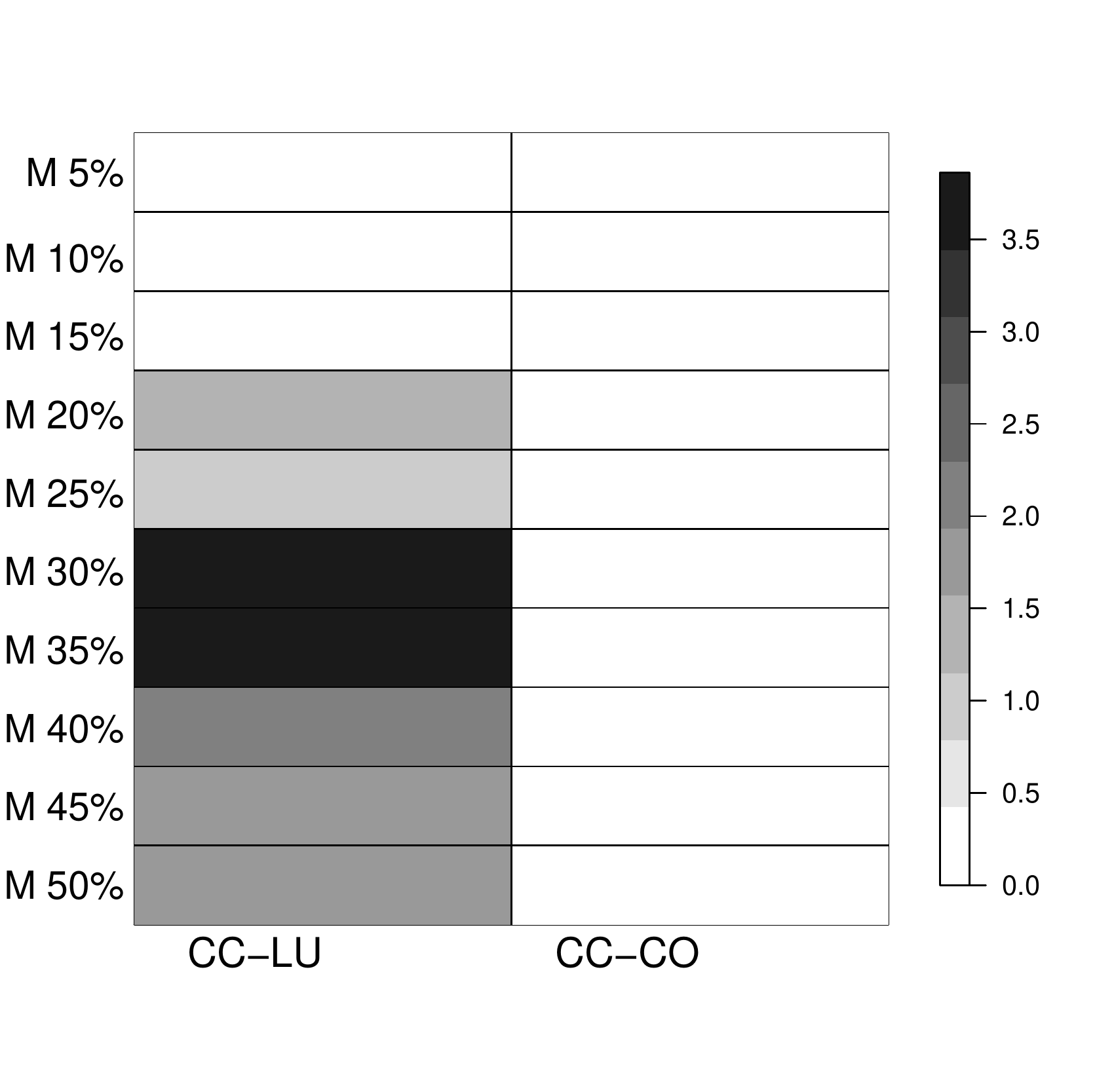}}
   \subfigure[Avg. Path Length]{\label{fig:LC-APL}\includegraphics[width=0.35\textwidth]{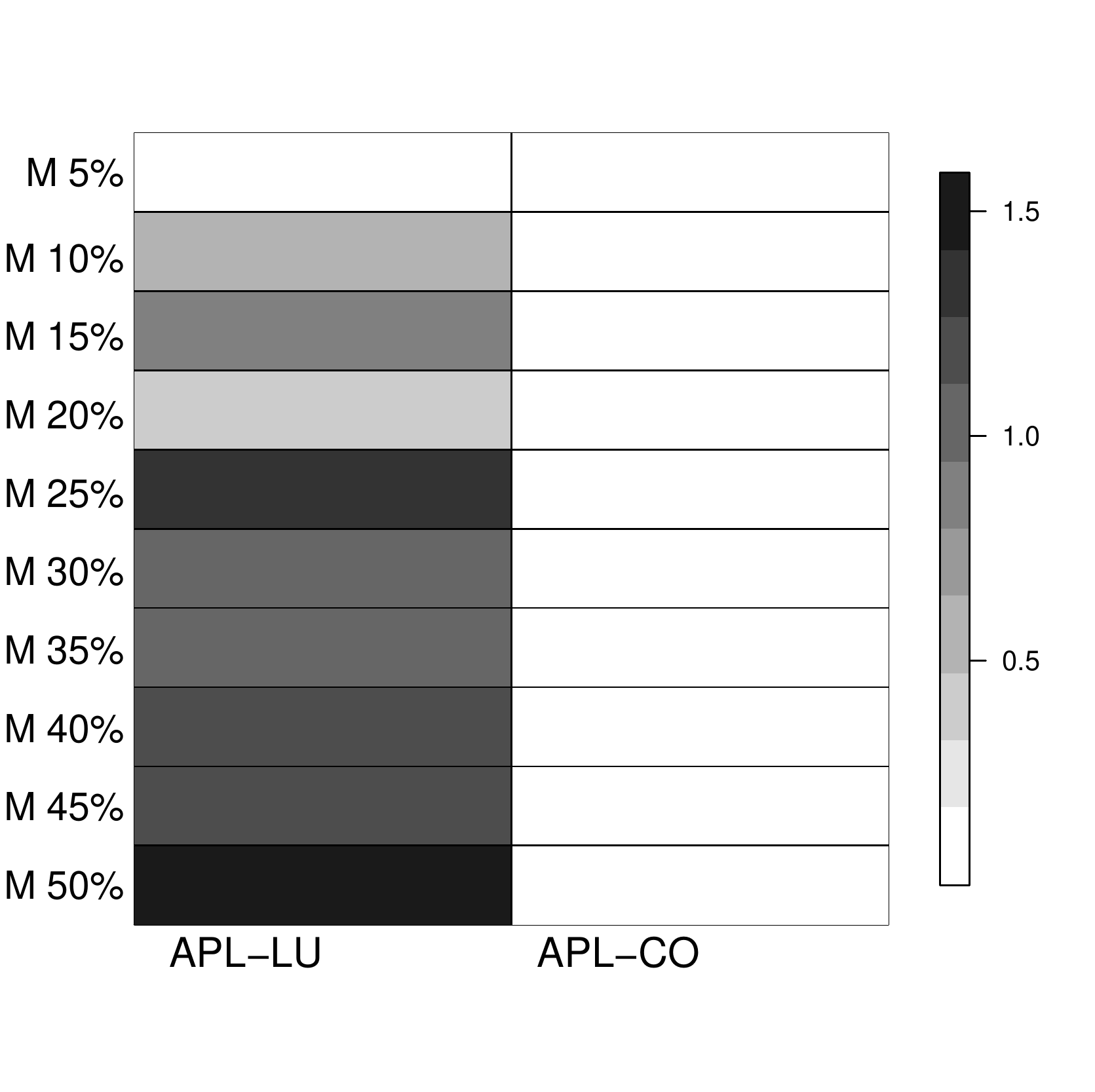}}
  \end{center}
  \caption{Comparison of Lunch and Coauthor layers.}
  \label{fig:EffOfSim}
\end{figure}

\section{Experiments}\label{sec:eval}

\begin{figure}[ht!]
  \begin{center}
    \subfigure[Diameter]{\label{fig:BB2LDia}\includegraphics[width=0.35\textwidth]{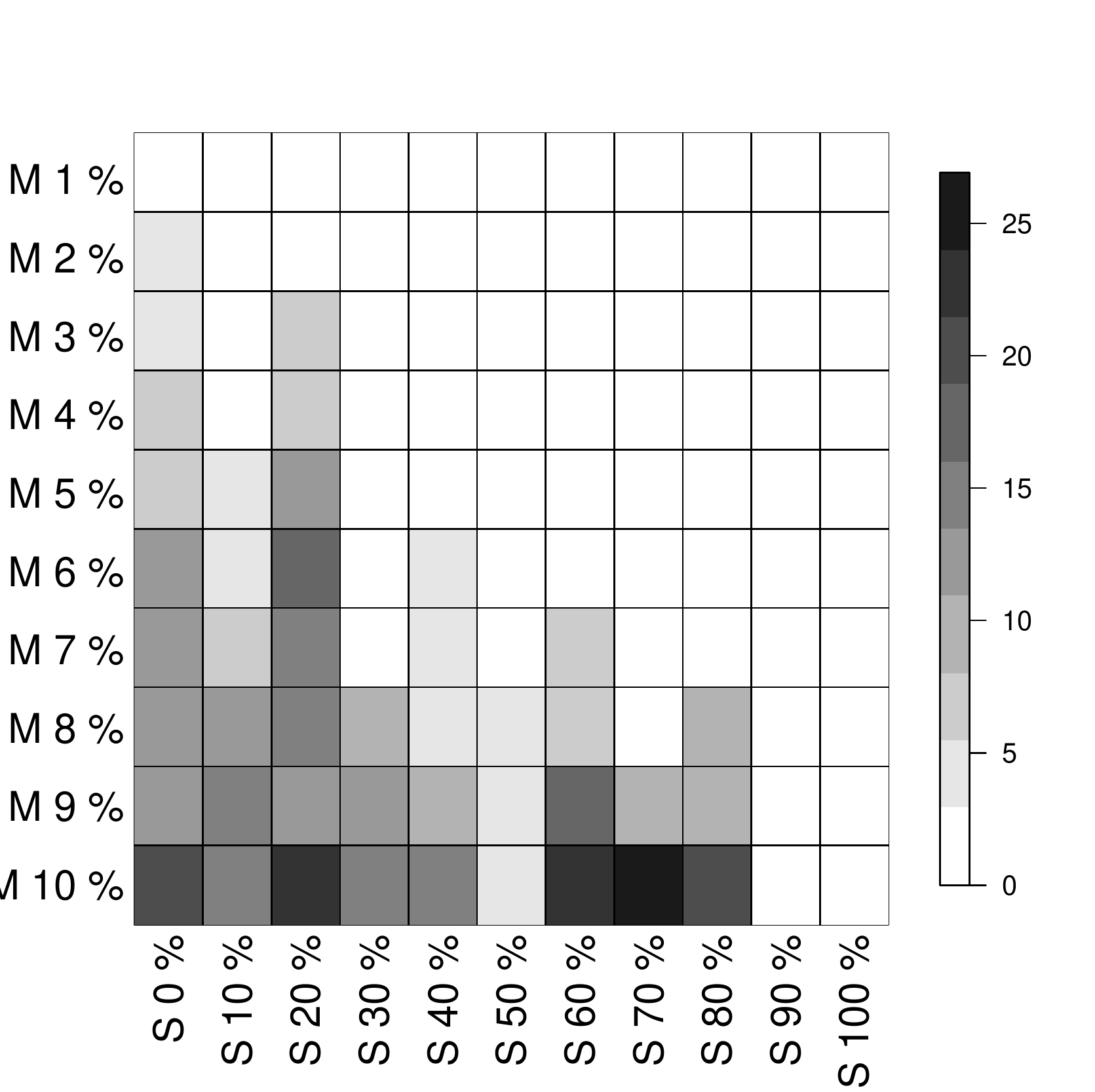}}
    \subfigure[Cluster Coeff.]{\label{fig:BB2LCC}\includegraphics[width=0.35\textwidth]{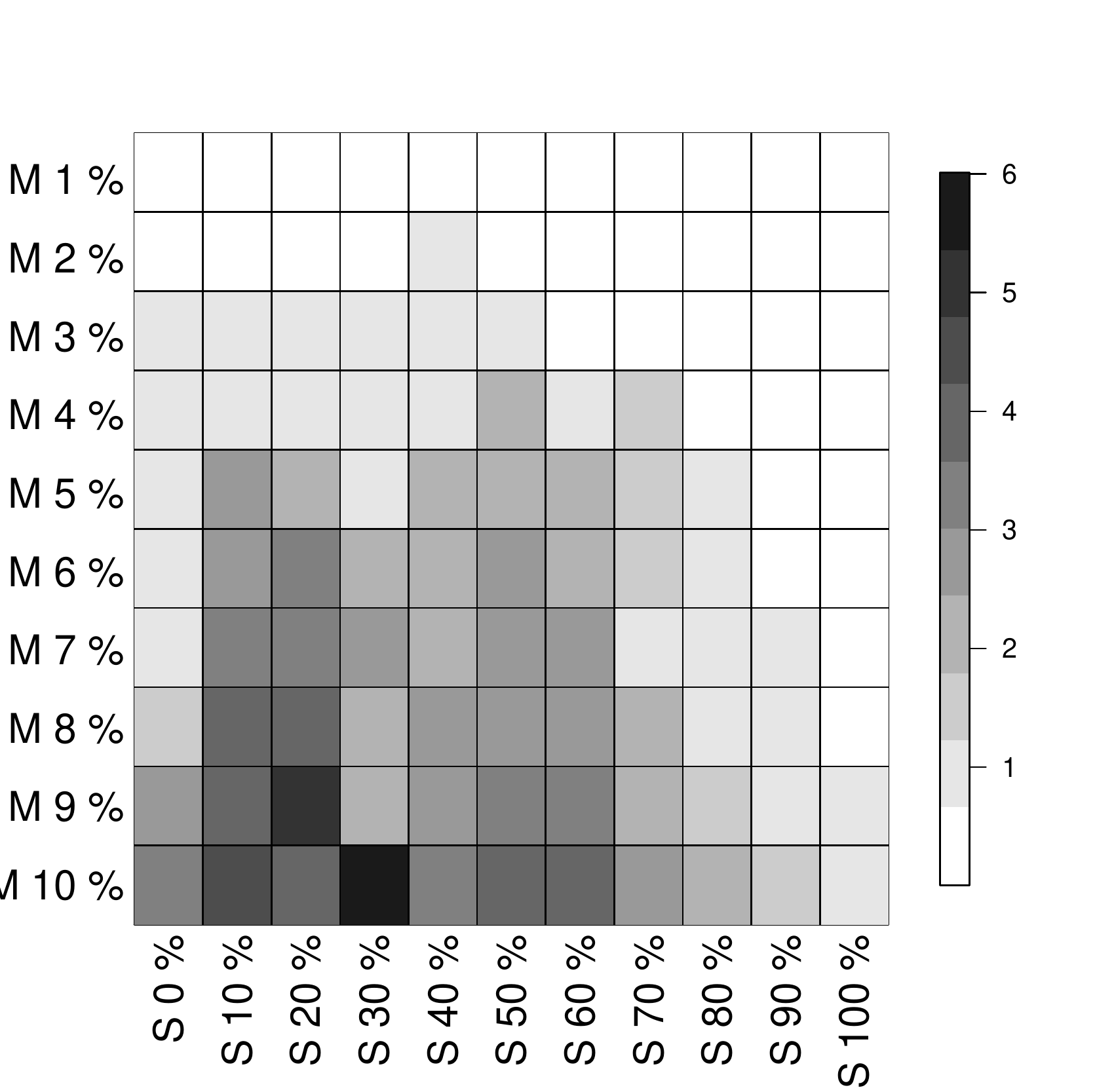}}
    \subfigure[Average Path Length]{\label{fig:BB2LApl}\includegraphics[width=0.35\textwidth]{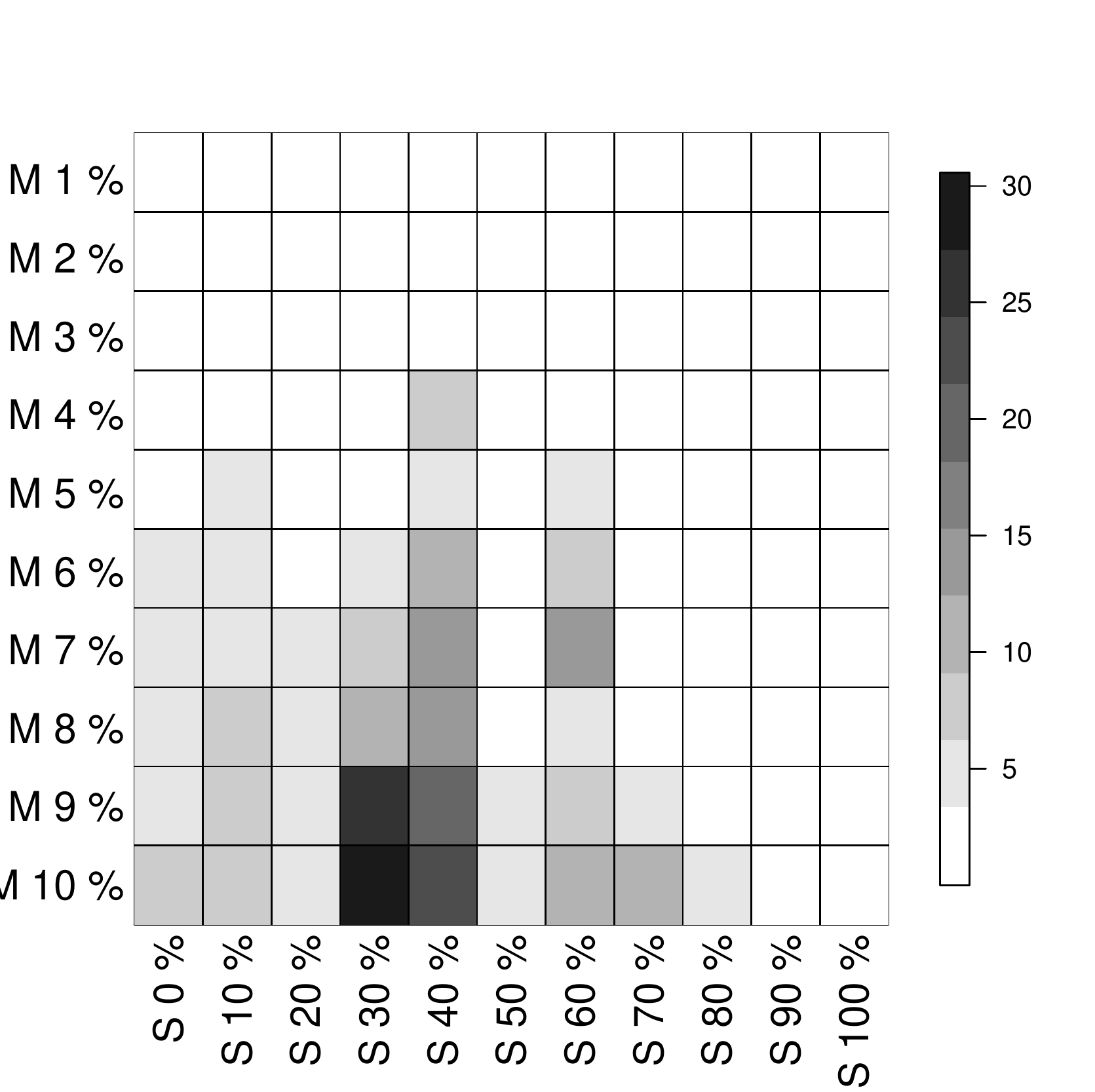}}
  \end{center}
  \caption{Effect of Missing Data in Multiple Layer Network.} \label{fig:MLVarySim}
\end{figure}

Now that we know how to interpret the numerical values resulting from our experiments (layer similarity, level of missing data and precision), we can apply our methodology to real and synthetic datasets.
\subsection{Datasets}
As the real and synthetic datasets have different properties, e.g., different sizes and different numbers of layers, we will use each of them to study the effect of missing data on specific properties.\\
\subsubsection{Real dataset }In the network of 61 employees of a University department (both faculty and administration) the following five kinds of interactions have been recorded: i) lunch, ii) work, iii) coauthor, iv) leisure, and v) Facebook. Thus, one can consider it as a five-layer network with 620 edges in the 5 layers. For more information about this dataset, that we call AUCS in the text, readers can refer to \cite{Magnani2013}.\\
\subsubsection{Synthetic dataset }As the real dataset doesnt have have varying properties thus, we created a synthetic dataset. The synthetic dataset consists of two-layer graphs with each layer having 10,000 nodes, where each layer graph is based on the Barabasi-Albert (BA) model \cite{barabassiModel} using the network generation framework presented in \cite{magnani2013formation}. We created 11 such multiplex networks with varying inter-layer similarity values, ranging from 0 to 100 with an interval of 10\%.


\subsection{Results}
In this section we present our experiment results for flatten and multiplex network properties.
We first present the results w.r.t effect of missing data on the flatten network.

\begin{figure*}[ht]
  \begin{center}
    \subfigure[Facebook]{\label{fig:FBXRelv}\includegraphics[width=0.3\textwidth]{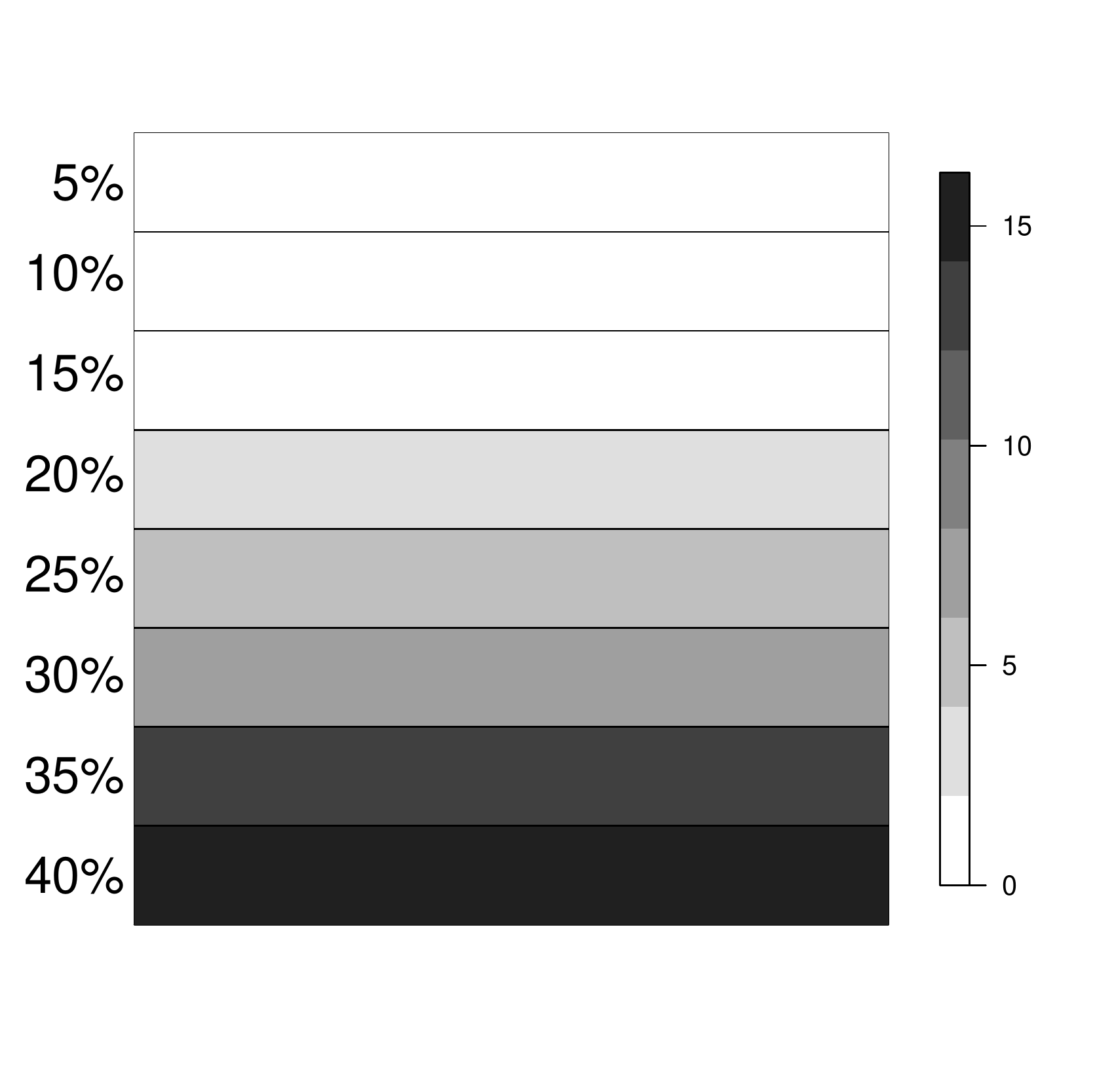}}
    \subfigure[Lunch]{\label{fig:LNXRelv}\includegraphics[width=0.3\textwidth]{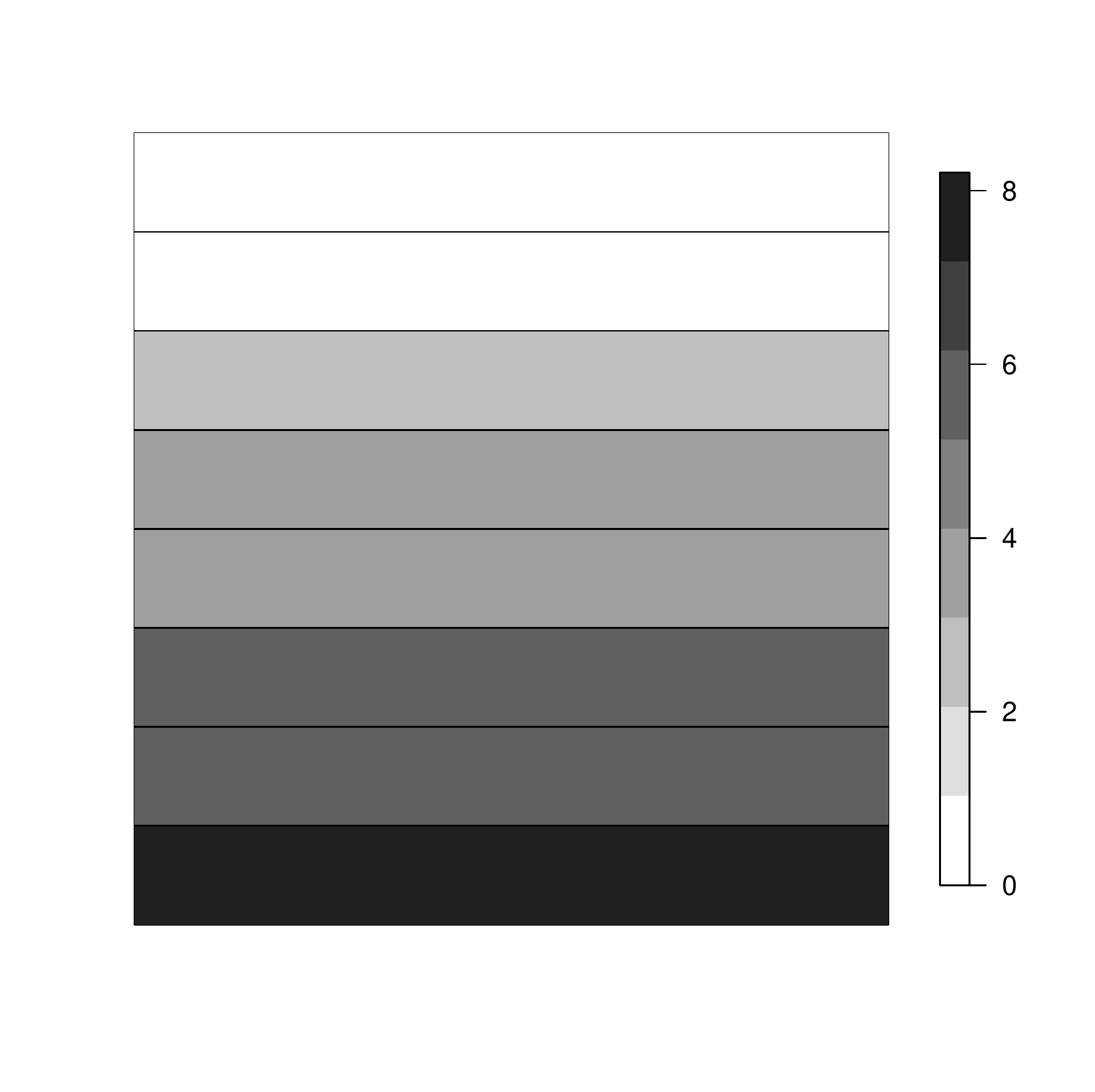}}
    \subfigure[Leisure]{\label{fig:LEXRelv}\includegraphics[width=0.3\textwidth]{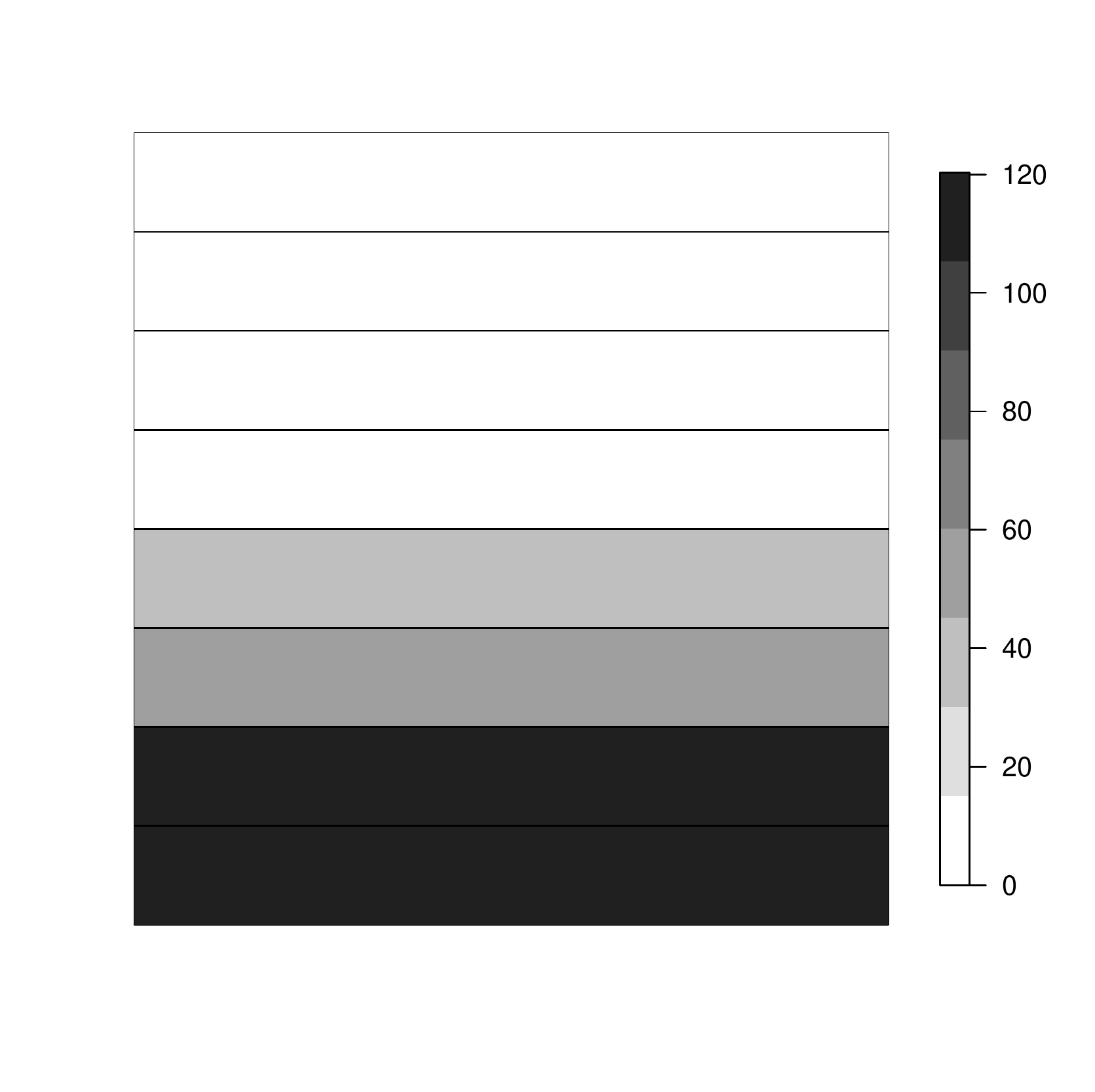}}
    \subfigure[Work]{\label{fig:WOXRelv}\includegraphics[width=0.3\textwidth]{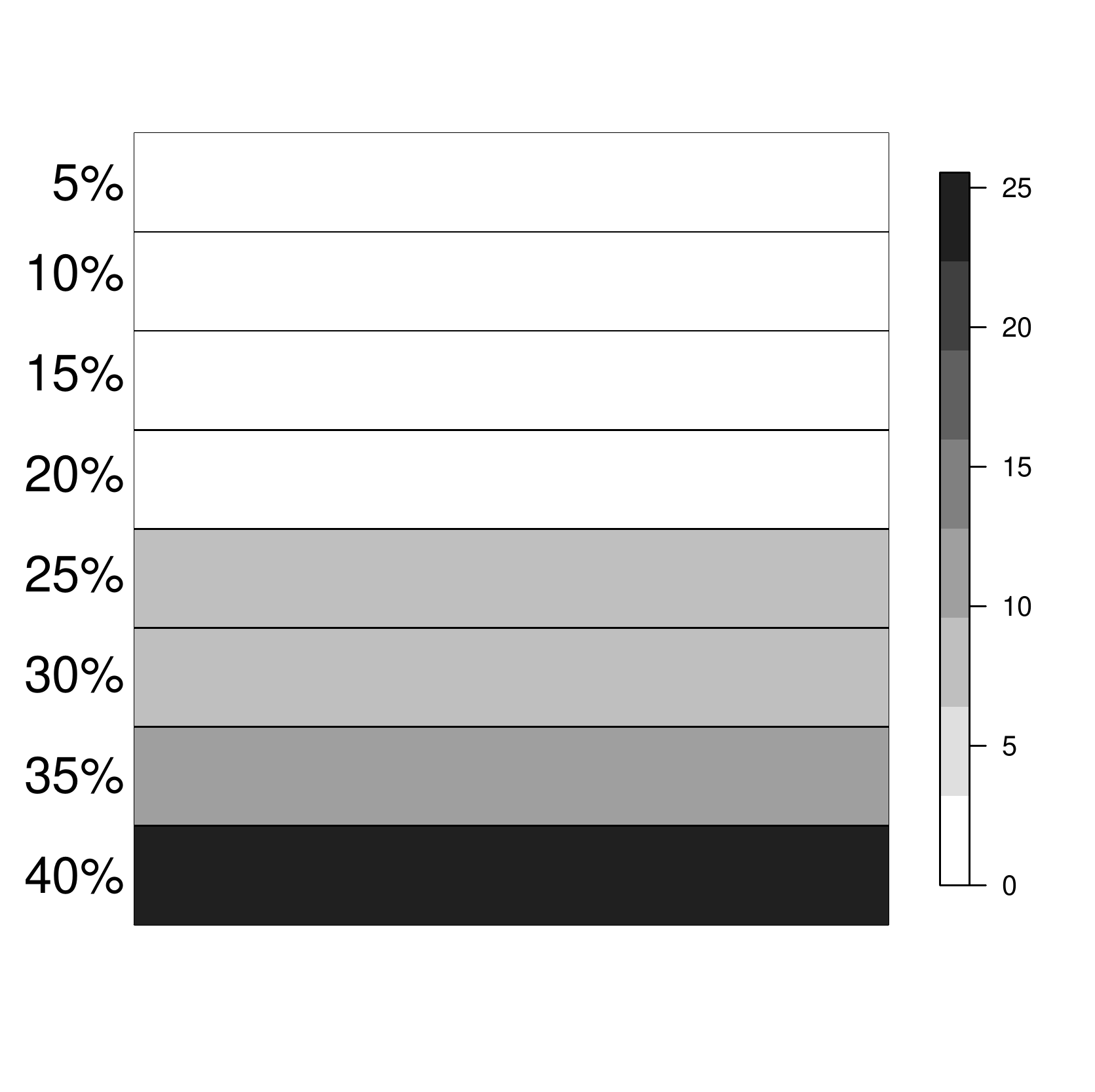}}
    \subfigure[Coauthor]{\label{fig:COXRelv}\includegraphics[width=0.3\textwidth]{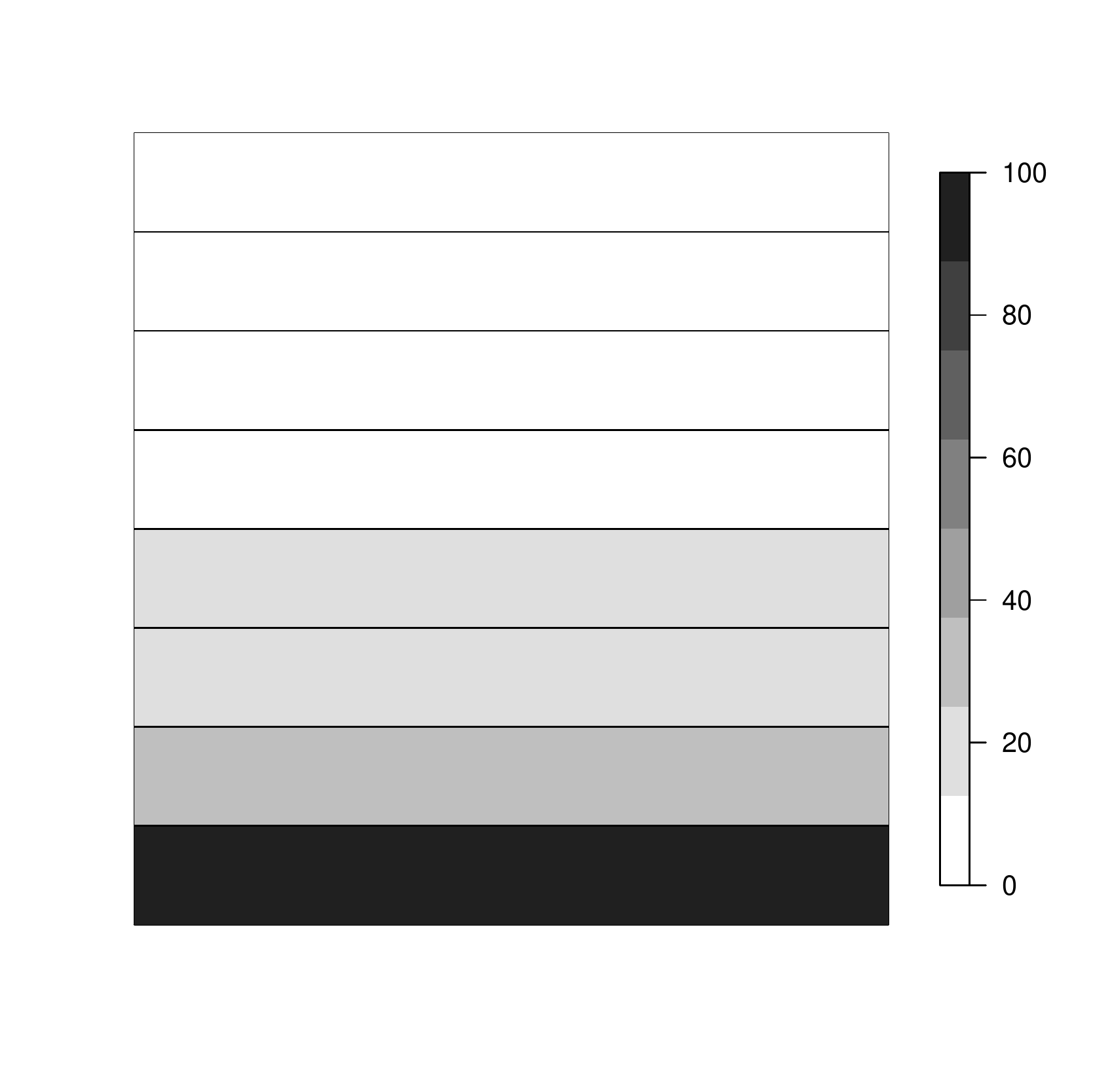}}
  \end{center}
  \caption{xRelevance on Five Layers.} \label{fig:XRelv}
\end{figure*}

\begin{figure*}[th!]
  \begin{center}
    \subfigure{\label{fig:center}\includegraphics[width=1\textwidth]{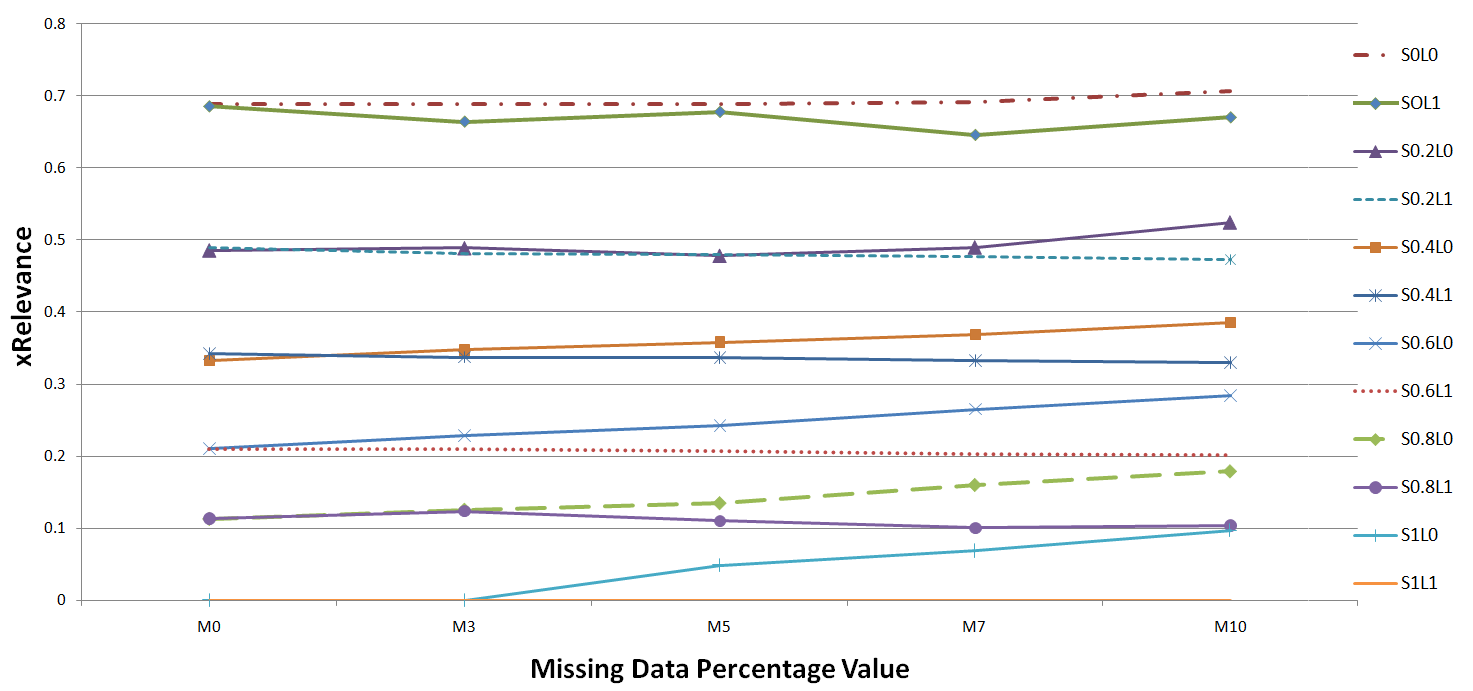}}
  \end{center}
  \caption{Effect of missing Data when data is removed from one of the layer.}
  \label{fig:SimMisDataLayers}
\end{figure*}

\subsubsection{Effect of similarity }
The similarity score of all the combined five layers of AUCS data is 0.009, with a similarity score of any two networks ranging from 0.339 to  0.058. To understand the effect of similarity, we selected two layers having the most and the least similarity values to rest of the layers. The $Lunch$ layer has the most similar value when compared to all the other layers,
and $Coauthor$ is the least similar layer to others. To see the effect of missing data, we start removing the edges from both these layers individually. Figure \ref{fig:EffOfSim} shows the effect of removal of edges from 5\% to 50\% in these two individual cases on Diameter (D), Clustering Coefficient (CC) and Average Path Length (APL) on Lunch (LU) and coauthor (CO) layers. Readers can refer to Table \ref{Tbl:VariablesTable} for all the short notations being used in the various experimental results. The figures are represented using monochromatic colour variations. Each box shows the \% variation of change of value w.r.t original value. The box of the colour for the corresponding missing \% and the respective property becomes more intense (thus darker in colour) with the increase in variation.

The lunch layer is the most similar layer to the overall network, compared to the other layers so one can expect that the effect of removal of the edges should be minimum. However, the absolute number of lunch edges (193) is almost 8 times higher than coauthorship (21). Thus, the removal of coauthor edges has not affected the properties as compared to removal of lunch layer. The removal of the edges has the least effect on the diameter. However, the removal of the 50\% of lunch edges has more negative impact on the other two structural properties of the graph in case of lunch layer. As the similarity score among layers is very less, thus, this real dataset is not ideal to see the effect of similarity values. To complement this, we present the next set of experiments on the synthetic dataset, which has a varying values of similarity.


\begin{table}[h!]
  \begin{center}
    \begin{tabular}{| c|  c |c |c|}
    \hline
    Layer Removal  & D         & CC                 & APL \\\hline
    \{CO,LU\}      & \{0,0\}   & \{0.25,11.55\}     &\{0.05,2.73\}  \\\hline
    \{FB,WO\}      & \{0,50\}  & \{3.40,7.28\}      & \{4.82,25.90\} \\\hline
    \{LU,WO\}      & \{0,25\}  & \{10.99,12.67\}    & \{2.65,21.58\} \\\hline
    \{WO,LE\}      & \{0,25\}  & \{0.83,1.08\}      & \{11.30,14.01\}  \\\hline
    \{LE,WO\}      & \{0,25\}  & \{1.12,1.08\}      & \{1.69,14.01\} \\\hline
    \end{tabular}
  \end{center}\label{Tbl:DiffComb}
   \caption{Effect of Layers Removal}
\end{table}


\subsubsection{Variation in similarity score of the layers }
When the similarity of the layers tends to 1 ($Sim(l_{i=1}^L)$ $\rightarrow$ 1), that is the networks in the individual layers are more similar to each other then the absence of nodes/edges in one layer does not affect much the flatten network properties. However, when the similarity score is low then the absence of nodes in one layer affects the flatten network's properties more adversely. This is established by our simulation on synthetic networks. Figure \ref{fig:MLVarySim} shows the results for three graph properties. In 11 multi layer networks, with varying similarity values (on the X-axis), we remove the edges from 1 to 10 \% (on the Y-axis) and calculate the three graph properties (D, CC, APL). The least common affected is the diameter among all the three. In all the three properties, we have not been able to find a consistent behavior with varying values. However, it is commonly observed that, with the decrease of similarity among layers and with an increase in missing values, more variation is observed.

\subsubsection{Effect of layer removal }
Table \ref{Tbl:DiffComb} shows the effect of layer removal on the flatten network properties on all the five layers, namely Facebook (FB), Lunch (LU), Work (WO), Leisure (LE) and Coauthor (CO). We take five individual cases, where we removed each layer and kept the four other layers to see the effect. To understand the effect of removal of a similar layer on the network, as a next step, we then remove the layer which is the most similar to remaining layers. We then notice the effect on the structural properties of the flatten network. We only removed up to two levels, as after that the graph becomes disconnected. Column 1 of the table shows the pair of layers removed one after the other and the corresponding column shows the values of percentage variations in D, CC, and APL properties after the removal of successive layers. A common observation is that there is a monotonic non-decreasing variation in the properties of the parameters.

\subsubsection{Multilayer network properties }
Figure \ref{fig:XRelv}, shows the xRelevance values for the five layers of the real dataset. We removed from 5\% to 40\% of the edges. In all the five layers there is a consistent change in xRelevance values, thus, signifying the importance of missing data. To see the importance of similarity on xRelevance, we performed experiments on synthetic dataset.
Figure \ref{fig:SimMisDataLayers} shows the impact of various similarities on xRelevance values on the synthetic dataset. The X-axis shows the missing data percentage and on the Y-axis the xRelevance values are plotted with different variations of similarity. A common observation in all the multilayer networks (with different similarity levels) is that with an increase in missing data, the gap between the xRelevance of layer 1 and layer 0 increases. 
These results show that there is a change in the xRelevance values on different layers and that as the data gets more and more incomplete on one of the layers, the relative importance of all the layers is affected. This is expected from the definition of xRelevance, and highlights how the presence of multiple layers introduces an additional level of complexity to the problem of missing data.

\begin{table}[!htbp]
  \small
	\caption{Legends used in various figures }
	
    \begin{tabular}{|p{2.4cm}|p{1.5cm}|p{2.8cm}|}
    \hline
    Type & Legend or Notation & Definition \\

	\hline
     \multirow{5}{*}{Network}
        & FB & Facebook  Layer.\\
		& LU & Lunch Layer.\\
        & WO & Work  Layer.\\
        & LE & Leisure  Layer.\\
        & CO & Coauthor  Layer.\\
  \hline
     \multirow{3}{*}{Network Properties}
        & D & Diameter.\\
		& CC & Cluster Coefficient.\\ 		
		& APL & Average Path Length.\\
				
\hline
     \multirow{2}{*}{Parameters}
 		& S & Similarity.\\
		& M & Missing.\\
	\hline
	\end{tabular}%
  \label{Tbl:VariablesTable}%
\end{table}%

\section{Conclusion and Future Work}\label{sec:concl}
Multiplex networks have been studied for a long time as they provide a more accurate representation of real world networks compared to single layer models. As in single layer networks, the problem of missing data in multiplexes is also naturally observed when the researchers collect network datasets. However, inherently because of complexity, missingness can impact multiplex network properties more adversely. In this preliminary study, we investigated the impact of missing data in multiplex networks on various network properties and provided a classification of the main causes of incompleteness. Furthermore, we performed experiments on real and synthetic datasets to understand the effect of missing data on various network properties.

\subsection{Future work}
We have a multidirectional future plan for this preliminary study:
\begin{enumerate}  
\item As an immediate next step is to perform experiments on to measure the effect of specific causes for missing data.

\item We will extend this study to perform experiments on other real datasets having a larger number of layers, with varying similarity values.
    
\item Another direction is to measure other multiplex network measures such as multidimensional betweenness centrality and multidimensional distance on different multiplex structures.

\end{enumerate}


\bibliographystyle{abbrv}
\bibliography{disco,additional_refs}

\begin{thebibliography}{10}

\bibitem{Bagrow2013}
J.~P. Bagrow, S.~Desu, M.~R. Frank, N.~Manukyan, L.~Mitchell, A.~Reagan, E.~E.
  Bloedorn, L.~B. Booker, L.~K. Branting, M.~J. Smith, B.~F. Tivnan, C.~M.
  Danforth, P.~S. Dodds, and J.~C. Bongard.
\newblock Shadow networks: Discovering hidden nodes with models of information
  flow.
\newblock {\em CoRR}, 2013.

\bibitem{barabassiModel}
A.-L. Barab\'{a}si and R.~Albert.
\newblock {Emergence of Scaling in Random Networks}.
\newblock {\em Science}, 286(5439):509--512, Oct. 1999.

\bibitem{Bearman2002}
M.~J. S.~K. Bearman, P.~S.
\newblock Chains of affection: The structure of adolescent romantic and sexual
  networks.
\newblock {\em ISERP Working Paper, Columbia University.}, 2002.

\bibitem{Berlingerio2012a}
M.~Berlingerio, M.~Coscia, F.~Giannotti, A.~Monreale, and D.~Pedreschi.
\newblock {Multidimensional networks: foundations of structural analysis}.
\newblock {\em World Wide Web}, Oct. 2012.

\bibitem{Bernard1984}
H.~R. Bernard, P.~Killworth, D.~Kronenfeld, and L.~Sailer.
\newblock {The Problem of Informant Accuracy: The Validity of Retrospective
  Data}.
\newblock {\em Annual Review of Anthropology}, 13(1):495--517, 1984.

\bibitem{Gjoka2011}
M.~Gjoka, M.~Kurant, C.~T. Butts, and A.~Markopoulou.
\newblock {Practical Recommendations on Crawling Online Social Networks}.
\newblock {\em IEEE J. Sel. Areas Commun. on Measurement of Internet
  Topologies}, 29(9):1872--1892, 2011.

\bibitem{Kivela2014}
M.~Kivela, A.~Arenas, M.~Barthelemy, J.~P. Gleeson, Y.~Moreno, and M.~A.
  Porter.
\newblock Multilayer networks.
\newblock {\em arXiv}, 2014.

\bibitem{Kossinets_SocNet2005}
G.~Kossinets.
\newblock {Effects of missing data in social networks}.
\newblock {\em Social Networks}, 28(3):247--268, July 2006.

\bibitem{lauman}
E.~O. Laumann, P.~V. Marsden, and D.~Prensky.
\newblock {\em {The boundary specification problem in network analysis..}},
  pages 18--34.
\newblock Sage Publications, London, 1983.

\bibitem{Leskovec2006}
J.~Leskovec and C.~Faloutsos.
\newblock {Sampling from large graphs}.
\newblock In {\em Proceedings of the ACM SIGKDD conference on Knowledge
  discovery and data mining}, pages 631--636, 2006.

\bibitem{LittleAndRubin}
Little and Rubin.
\newblock Statistical analysis of missing data.

\bibitem{Magnani2013}
M.~Magnani, B.~Micenkov{\'a}, and L.~Rossi.
\newblock Combinatorial analysis of multiple networks.
\newblock {\em CoRR}, abs/1303.4986, 2013.

\bibitem{MagnaniJDIQ10}
M.~Magnani and D.~Montesi.
\newblock {A Survey on Uncertainty Management in Data Integration}.
\newblock {\em ACM Journal of Data and Information Quality}, 2010.

\bibitem{magnani2013formation}
M.~Magnani and L.~Rossi.
\newblock {Formation of multiple networks}.
\newblock In {\em Social Computing, Behavioral-Cultural Modeling and
  Prediction}, pages 257--264. Springer Berlin Heidelberg, 2013.

\bibitem{MatteoMuldiDimNw}
M.~A. R. G. G.~F. Magnani, M.
\newblock On multidimensional network measures.
\newblock In {\em Italian Conference on Sistemi Evoluti per le Basi di Dati
  (SEBD)}, 2013.

\bibitem{Marsden1990}
P.~V. Marsden.
\newblock {\em {Network data and measurement.s}}, pages 435--463.
\newblock 1990.

\bibitem{Roethlisberger1939}
D.~W.~J. Roethlisberger, F.~J.
\newblock Management and the worker.
\newblock {\em Harvard University Press, Cambridge, MA}, 1939.

\bibitem{Rumsey1993}
D.~J. Rumsey.
\newblock Nonresponse models for social network stochastic processes (markov
  chains).
\newblock 1993.

\bibitem{Sadikov2011}
E.~Sadikov, M.~Medina, J.~Leskovec, and H.~Garcia-Molina.
\newblock Correcting for missing data in information cascades.
\newblock In {\em Proceedings of the Fourth ACM International Conference on Web
  Search and Data Mining}, WSDM '11, pages 55--64, New York, NY, USA, 2011.
  ACM.

\bibitem{Salehi2013}
M.~Salehi and H.~R. Rabiee.
\newblock A measurement framework for directed networks.
\newblock {\em IEEE Journal on Selected Areas in Communications (JSAC): Special
  Issue on Network Science}, 31(6):1007--1016, 2013.

\bibitem{Travers69anexperimental}
J.~Travers and S.~Milgram.
\newblock An experimental study of the small world problem.
\newblock {\em Psychology Today}, 2:60--67, 1967.

\end{thebibliography}
\label{sec:References}
\end{document}